\documentclass[aps,prd,reprint,amsmath,amssymb]{revtex4-1}
\usepackage{graphicx}
\usepackage{dcolumn}
\usepackage{braket}
\usepackage{bm}
\usepackage{color}

\begin{document}

%\preprint{}

\title{Vacuum Persistence in Fierz-Pauli Theory on a Curved Background}

\author{Sungmin Hwang}
 \email{sungmin.hwang@tum.de}
 \affiliation{Physik-Department, Technische Universit{\"a}t M{\"u}chen, James-Franck-Str.1, D-85747 Garching, Germany}
 \affiliation{Max-Planck-Institut f{\"u}r Physik, F{\"o}hringer Ring 6
D-80805 M{\"u}chen, Germany}
\author{Dennis Schimmel}
 \email{dennis.schimmel@physik.lmu.de}
 \affiliation{Arnold Sommerfeld Center for Theoretical Physics, Ludwig-Maximilians-Universit{\"a}t M{\"u}chen, Theresienstr. 37, D-80333 M{\"u}chen, Germany}
%\thanks{}


\date{\today}

\begin{abstract}
By explicitly constructing the Hilbert space, Higuchi showed that there is a lower bound on the mass of a minimally-coupled free spin-2 field on a curved background \cite{HiguchiBound}. Using the vacuum persistence amplitude, we show that this bound is modified by taking into account additional terms not prohibited by symmetry in the case of a maximally symmetric spacetime. This result can further be generalized to the maximally symmetric space case, such as the FRW universe, and its corresponding bound of the deformation parameter is discussed. 

%It is well known from the past that a lower bound of the deformation parameter to the Einstein's general theory of relativity arises in a non-interacting and minimally-coupled massive spin-2 field theory, with a maximally symmetric spacetime as the given classical background (de Sitter spacetime in particular).  We show that the known bound is modified when all possible terms, which are allowed by symmetry of the theory, are considered, including the non-minimally coupled term, and argue that background spacetime dependent lower bound to the mass parameter is absent. This result can further be generalized to the maximally symmetric space case, such as the FRW universe, and its corresponding bound of the deforamation parameter is discussed. 
\end{abstract}

%\keywords{}

\maketitle

\section{Introduction and Outline}
The cosmological constant problem is one of the most outstanding challenges in theoretical physics. The observed value of the vacuum energy density from the redshifts of distant galaxies and supernovae is bounded from above by $10^{-48} \text{GeV}^{4}$ (by setting $\hbar = c = 1$) \cite{DarkE1, DarkE2}, while the change in the vacuum energy density, which is expected to occur during electroweak symmetry breaking, is $\sim 10^{8} \text{GeV}^{4}$ ; raising the scale via renormalization group flow worsens the discrepancy such that the predicted change in the vacuum energy density becomes $\sim 10^{72} \text{GeV}^{4}$ at the Planck scale \cite{WeinbergCC}. A striking amount of fine-tuning is required in order for these scales to be matched with observations. Thus, this challenge calls for any kind of a self-consistent UV completion of quantum gravity or some modifications to either gravity or quantum field theory. In this paper, we take modification of gravity (Einstein's general relavity) at linear level (namely the \emph{Fierz-Pauli theory}), which attempts to resolve this issue. A deformation parameter (i.e., mass term) to general relativity changes the gravitational potential to the Yukawa-type, such that its strength decreases exponentially at large distances $r\sim 1/m$, where $m$ is the deformation parameter, and thereby suggests a possible explanation to the tiny observed value of the cosmological constant in the universe \cite{GiaStefan, Zee, WeinbergQFT}.\\

Since we are interested in phyiscs in the \emph{far} infrared regime and vertices in general relativity (and modifications thereof), scale as $(k^{2}/M^{2}_{\text{Pl}})^{n}$ \cite{Zee}, it is sufficient at first to consider the linearized theory and absorb the effects of the interaction in a renormalization of the parameters.  This approximation is implicitly encoded from the long-ranged estimate from the previous paragraph, which can be translated into $E \sim 1/r \sim m$, where the deformation parameter $m$ (the graviton mass) is a tiny value from the observational constraint comparing to the Planck scale \cite{Gmass, Gmass2, Gmass3}. A problem arises when one takes the massless limit of the linearized theory, called van Dam-Veltman-Zakarov (vDVZ) discontinuity \cite{vDVZ1, vDVZ2}, which causes the discrepancy in certain experimental constraints (deflection angle of a light beginning from stars, for expample). Since general relativity is a non-linear theory, including the higher order terms can smoothly recover the massless limit of the spin-2 theory by the Vainshtein mechanism \cite{Vainshtein}; the massive spin-2 theory is only applicable outside the \emph{Vainshtein radius}, which is defined as $r_v = (G_N M/m^{4})^{1/5}$, where $M$ is the mass of static source and $G_N$ is the Newton's contant. That is, the Vainshtein radius sets up a cutoff of the theory, and when the massless limit of the graviton is taken, the Vainshtein radius becomes infinite, thereby the regime of massless spin-2 theory is recovered. Furthermore, there are some cases in which the massive spin-2 theory does not suffer from the vDVZ discontinuity, such as on the negatively curved maximally symmetric spacetime \cite{Neupane}. 
\\

When the degravitation effect on a generic curved background is considered, such as the theory of massive gravity on a curved background at the far infrared regime, one has to pay enough attention to its construction, especially the ``covariantization" procedure for the Pauli-Fierz theory \cite{FPmass}. The only safe way of covariantizing the Fierz-Pauli theory is to include all possible terms that are not prohibited by symmetries and keep couplings to terms vanishing in a flat (Minkowski) background as free parameters. This can only be avoided if the full model including its interactions with background geometry is known, which is not the case here. Thus, one should consider not only the Fierz-Pauli mass term \cite{FPmass}, but also terms coupling the field to the background curvature.  These terms  modify theoretical bounds on the allowed values of the mass. A possible way of taking this into account is to compute the vacuum persistence amplitude (VPA), which is a diagnostic tool for testing unitarity of the theory. If the quantum theory at the leading order violates the unitarity condition, it is unlikely that the full theory would be able to cure the violation as far as a legitimate perturbation scheme is concerned. We pay attention to de Sitter and Friedmann-Robertson-Walker (FRW) spacetimes as the given classical backgrounds in this discussion, for those are good zeroth order approximations to the universe we observe today. \\

%Proper covariantization to the linearized massive spin-2 theory in a classcal background is required once its scale of approximation is set up. Not only is the (Fierz-Pauli) mass term to be added \cite{FPmass}, but a non-minimal coupling between the background spacetime curvature and the graviton field is also to be included since one has to consider all possible terms of quadratic order in the corresponding field in accordance with the effective field theory language. It is a common misconception from the past that the non-minimal coupling may be dropped, from which it led to the consensus that there exsists a non-trivial background curvature dependent lower bound of the graviton mass when the background is specified to the de Sitter spacetime \cite{HiguchiBound}; the known mass bound is modifed once the non-minimal coupling is taken into account. This modification comes out naturally when one uses the vacuum persistence amplitude (VPA) as a diagnostic tool for testing unitarity of the theory. 

The program of testing unitarity proceeds as follows. We construct the Lagrangian of the linearized massive spin-2 field on a classical background in section II. Riemann normal coordinates are used to obtain a consistent approximation scheme. Results are, within this scheme, computed up to second adiabatic order. The general expression of the VPA is given in section III, and by using the Green's function derived in the previous section along with inserting a localized static external source, we evaluate the unitarity of the model. Imposing unitarity gives rise to a new feature in the lower bound of the deformation parameter, and its implications are discussed in section IV. Further analyses and implications to another maximally symmetric spacetime with negative curvature (anti-de Sitter) and maximally symmetric space (FRW universe) are discussed in the same section. Lastly, more detailed calculations of the Green's function are presented in the appendix section.

%Further reasons why Higuchi's bound is weird? 

%The Riemann normal coordinate is employed to restrict the region of the adiabatic approximation when the background is specified to the de Sitter spacetime, and two-point correlation function is computed up to the second adiabtic order by defining a proper vacuum state in the restricted region. 

\section{Linearized Massive Gravity in a Classical Background}
\subsection{Lagrangian of the Linearized Massive Gravity}
The dynamics of the massive spin-2 theory on a curved background is obtained from the Fierz-Pauli action \cite{FPmass} when we denote the metric of the background by $\bar{g}(x)$ and fluctuation around it with $h_{\mu\nu}(x)$. The cosmological constant $\Lambda$ is included as an explicit source field, and the full action is schematically given, after canonically normalizing the fluctuation field \cite{Aragone}, by 
\begin{eqnarray}\label{full action}
S = S_{0} + S_{m} + S[h^{3}]
\end{eqnarray}
in which $S_{0}$ contains all quadratic terms of the massless spin-2 field 
\begin{eqnarray}\label{linear action}
S_{0} &=& \int d^{4}x\sqrt{-\bar{g}} \Bigl[\frac{1}{2}\nabla^{\sigma}h^{\alpha\rho}\nabla_{\alpha}h_{\rho\sigma} - \frac{1}{4}\nabla^{\sigma}h^{\alpha\rho}\nabla_{\sigma}h_{\alpha\rho} \nonumber \\
&& + \frac{1}{4}(\nabla^{\rho}h - 2\nabla^{\alpha}h^{\rho}_{\alpha})\nabla_{\rho}h  +  (h^{\alpha\sigma}h^{\beta}_{\sigma} - \frac{1}{2}hh^{\alpha\beta})\bar{\mathcal{R}}_{\alpha\beta} \nonumber \\
&& - \frac{1}{4}(h^{\alpha\beta}h_{\alpha\beta} - \frac{1}{2}h^{2})(\bar{\mathcal{R}} - 2\Lambda) - \frac{1}{4}\xi\bar{\mathcal{R}}(h_{\alpha\rho}h^{\alpha\rho} - h^{2}) \Bigr] \nonumber \\
&\equiv & \int d^{4}x\sqrt{-\bar{g}}\mathcal{L}_{0}
\end{eqnarray}
where $h = h^{\mu}_{\mu} \equiv \bar{g}_{\mu\nu}h^{\mu\nu}$. The bar notation on top of the Ricci tensor and scalar implies that these quantities are evaluated in terms of the background metric $\bar{g}$. Note that the last term which carries the coupling constant $\xi$ indicates the coupling between the fluctuation field and the background. Since $\xi$ is a free parameter not affected by any symmetries, we expect it to be of order one. There are several additional ways to couple the Ricci (or Riemann) tensor to the fluctuation field, but all of them are reduced down to this expression for a maximally symmetric background. $S_{m}$ is the mass contribution to the action whose expression is given by \cite{FPmass}
\begin{eqnarray}
S_{m} &\equiv &\int d^{4}x\sqrt{-\bar{g}}\mathcal{L}_{m} \nonumber \\
&=& -\frac{1}{2}\int d^{4}x\sqrt{-\bar{g}}[m^{2}(h_{\mu\nu}h^{\mu\nu} - h^{2})].
\end{eqnarray}
In principle, one has to include this deformation term with arbitrary relative coefficients for $h_{\mu\nu}h^{\mu\nu}$ and $h^{2}$, but the only legitimate expression which ensures stability of the theory (i.e., absence of ghost) is the given expression. Inclusion of the deformation term, along with the non-minimal coupling term, seemingly breaks the gauge symmetry of the theory (general covarince), but this seemingly broken symmetry can be restored by the St{\"u}ckelberg formalism just as shown in the massive spin-1 case of the electromagnetism \cite{Stuckelberg, Alex}. That is, there exists a new set of symmetries in the linearized massive spin-2 theory by introducing ``St{\"u}ckelberg vector and scalar fields". The rest of the terms in $h$ with cubic orders or higher are included in $S[h^{3}]$, and our interest is only on the first two terms on the right hand side of the eq. (\ref{full action}). \\

Historically, the action of a massive spin-2 field was obtained from Einstein's general relativity. The idea was to simply take the Einstein-Hilbert action and expand it to leading order in small fluctuations around a given background (which in turn solves the full non-linear problem). More explicity, the action of the massless spin-2 theory on a classical background is derived by the perturbative expansion of general relativity, $g_{\mu\nu}(x) = \bar{g}_{\mu\nu}(x) + h_{\mu\nu}(x)$ along with taking canonical normalization of the field. Including the terms up to second order in the metric fluctuation $h_{\mu\nu}$ yields the minimally-coupled ($\xi=0$) linearized massless spin-2 theory \cite{Wald}. Then, the Fierz-Pauli mass term was added by hand to the leading order in fluctuations. From the context of effective field theories, this procedure leaves a lot to be desired, as this approach does not generate the term proportional to $\xi$ in eq. (\ref{linear action}). While the result is a consistent theory (at least to leading order) in its own, it is not the most general action compatible with all symmetries given. Instead, if Einstein's general relativity is to be modified, all allowed terms should be taken into account. \\

Hence, the full expression of the linearized action for the massive spin-2 theory at the weak-field limit, with an external source added, is summarized as
\begin{equation}\label{action}
S[h; T] = \int d^{4}x\sqrt{-\bar{g}}[\mathcal{L}_{0} + \mathcal{L}_{m} + h_{\mu\nu}T^{\mu\nu}],
\end{equation}
and the equations of motion are derived by taking a variation of the action with respect to the fluctuation field
\begin{eqnarray}
\frac{1}{\sqrt{-\bar{g}}}\frac{\delta S}{\delta h^{\alpha\rho}}  &=&  -(\mathcal{K}_{m=0}h)_{\alpha\rho} - m^{2}(h_{\alpha\rho} - \bar{g}_{\alpha\rho}h) + T_{\alpha\rho} \nonumber \\
&=& 0
\end{eqnarray}
in which $\mathcal{K}_{m=0}$ is a sum of the kinetic operator and curvature contribution to the massless part. Since the massless spin-2 theoy has two transversal modes, $\mathcal{K}_{m=0}$ is a transverse projection operator, and the Bianchi identiy (by taking a convariant derivative on the equations of motion) imposes four constraints $\nabla^{\alpha}h_{\alpha\rho} = \nabla_{\rho}h$ with $\nabla_{\rho}$ being a covariant derivative in the curved background. A commutation relation in the covariant derivatives has to be found in order to use these constraints, which is 
\begin{eqnarray}
[\nabla_{\sigma}, \nabla_{\alpha}]h^{\sigma}_{\rho} &=& \bar{\mathcal{R}}_{\lambda\alpha}h^{\lambda}_{\rho} - \bar{\mathcal{R}}_{\lambda\rho\sigma\alpha}h^{\lambda\sigma} \\
\nabla_{\sigma}\nabla_{\alpha}h^{\sigma}_{\rho} &=& \nabla_{\alpha}\nabla_{\sigma}h^{\sigma}_{\rho} - \frac{\Lambda}{3}(\bar{g}_{\alpha\rho}h - 4h_{\alpha\rho})
\end{eqnarray}
where the second line is given by specifying de Sitter spacetime as the background metric. The equations of motion with five independent components are derived by using this commutation relation and the constraints given above, and taking the trace out (for $h$ is a non-dynamical degree of freedom)
\begin{eqnarray}
&&(-\Box^{c} + \mu^{2})h_{\alpha\rho} + \frac{2\Lambda}{3}h_{\alpha\rho}  \nonumber \\
&&= T_{\alpha\rho} + \frac{1}{3\mu^{2} - 2\Lambda}\Bigl[\nabla_{\rho}\nabla_{\alpha}T - \mu^{2}\bar{g}_{\alpha\rho}T + \frac{\Lambda}{3}\bar{g}_{\alpha\rho}T\Bigr]
\end{eqnarray}
in which $\mu^{2} \equiv m^{2} + 4\xi\Lambda$ and $\Box^{c} \equiv \nabla^{\mu}\nabla_{\mu}$. 
Let us denote this relation by $(\mathcal{O}h)_{\alpha\rho} \equiv (\Pi T)_{\alpha\rho}$. It is schematically solved by $h = G*T$ ($*$ is the convolution) where $G$ is the Green's function. Since it is difficult to construct a Green's function globally, we compute it only in its adiabatic expansion. To do so, first the correct vacuum (another difficult issue on general backgrounds) has to be identified. This can be done consistently in an adiabatic expansion.

\subsection{Adiabatic Vacuum and Green's functions}
The concept of a ``unique'' vacuum state on a generic classical background is ambiguous due to general covariance  \cite{ParkerToms, Birell, Mukhanov}. Quantum field theory in flat spacetime is constructed upon the ``unanimous'' notion of vacuum: A physically unambiguous vacuum state with respect to one inertial reference frame is also a vacuum in the other inertial reference frames because there exists a preferred global timelike Killing vector field $\partial_{t}$ in the Poinc{\'a}re group, which is orthogonal to the spacelike hypersurfaces. That is, when the mode expansion is taken on the solution to the equations of motion for a quantum field in flat spacetime, a basis is chosen such that the solution is unique up to a phase because of the timelike Killing vector field. As the symmetry of the background is changed from Poincar{\'e} invariance to general covariance, however, we do not have such a preferred Killing vector, for the manifold is coordinate independent; there exists several different Killing equations \cite{Birell, ParkerToms} so that the notion of ``preferred'' basis is no longer available on a curved background. For instance, when asymptotic ``in" (for $t\rightarrow -\infty$) and ``out" (for $t\rightarrow +\infty$) states are defined in the expanding universe, the basis of mode expansions for the asymptotic states are non-trivially related by the Bogoliubov transformation \cite{Birell}. The vacuum expectation value of the particle number operator is not necessarily equal to zero if the ``in" state number operator is evaluated with respect to the ``out" vacuum and vice versa, thereby a unique particle interpretation is lacking \cite{ParkerToms, Birell, Mukhanov}. 
\\

However, an approximate vacuum state can still be defined by imposing certain constraints on the background. It is well-known for a slowly varying (or expanding) FRW universe that one can solve the equations motion by WKB approximation \cite{ParkerToms, Mukhanov}. In the case of a massive scalar field theory in the FRW universe, the slowly varying background implies that the period of the harmonic oscillator is much smaller than an adiabatic parameter, which we call $T$, and this parameter can be chosen as the Hubble time $H^{-1}$. In other words, we take the WKB approximation to the solution based on an assumption that the wavelength of the scalar particle is much smaller than the curvature scale of the universe, $\lambda_{k} \ll H^{-1}$, and thereby the asymptotic ``in" and ``out" states are defined just as in the flat spacetime. The approximation scheme in this paper is up to second adiabatic order, and its corresponding vacuum state is the second adiabatic one $\ket{0_{(2)}}$ \cite{ParkerToms}. \\

This approximation scheme becomes more explicit by a particular choice of the coordinate system of the background. We employ Riemann normal coordinate (or a local inertial frame), centered at $x'$. The metric is then given by \cite{ParkerToms}
\begin{eqnarray}
\bar{g}_{\mu\nu}(x; x') &=& \eta_{\mu\nu} - \frac{1}{3}\bar{\mathcal{R}}_{\mu\alpha\nu\rho}(x')y^{\alpha}y^{\rho} \nonumber \\
&& + \frac{1}{3!}\bar{\mathcal{R}}_{\mu\alpha\nu\rho; \sigma}(x')y^{\alpha}y^{\rho}y^{\sigma} + \cdots
\end{eqnarray}
where $y = x - x'$ and the spacetime indices are raised and lowered by the flat metric $\eta_{\mu\nu}$. We can rewrite the expansion scheme shown above as $y\ll H^{-1} \sim \Lambda^{-1/2}$ (or $\Lambda y^{2} \ll 1$) if the background is chosen to be the maximally symmetric spacetime. In other words, we consider a small region around an arbitrary origin $x'$. This region is much smaller than the surface of Hubble sphere with radius $r_H\sim H^{-1}$, and thus is well approximated by small corrections to Minkowski. The vacuum structure, in particular, is very similar to that of fluctuations around Minkowski spacetime. Since the curvature scale $\Lambda$ is a fixed value in the maximally symmetric spacetime, the approximation up to second adiabatic order means that only the terms up to the quadratic order in $y$ are to be included in the action. The de Sitter metric and its inverse metric in the normal neighberhood are given below
\begin{eqnarray}
\bar{g}_{\alpha\beta}(y) &=& \eta_{\alpha\beta} - \frac{\Lambda}{9}\eta_{\alpha\beta}y^{2} + \frac{\Lambda}{9}y_{\alpha}y_{\beta} +\mathcal{O}(y^{3})\nonumber \\
\bar{g}^{\tau\gamma}(y) &=& \eta^{\tau\gamma} + \frac{\Lambda}{9}\eta^{\tau\gamma}y^{2} - \frac{\Lambda}{9}y^{\tau}y^{\gamma} + \mathcal{O}(y^{3})
\end{eqnarray}
from which the Christoffel symbol and Riemann tensor are computed accordingly. \\

With this power counting scheme, we can expand the left hand side of the equation of motion up to the second adiabatic order as follows
\begin{eqnarray}\label{left}
(\mathcal{O}h)_{\alpha\rho} &\equiv & (-\Box^{c} + \mu^{2} + \frac{2\Lambda}{3})h_{\alpha\rho} \nonumber \\
&=& (-\Box + \mu^{2})h_{\alpha\rho} + \frac{2\Lambda}{9}y^{\lambda}\partial_{\lambda}h_{\alpha\rho} - \frac{2\Lambda}{9}y_{(\alpha}\partial^{\lambda}h_{\rho)\lambda} \nonumber \\
&& + \frac{4\Lambda}{9}y^{\lambda}\partial_{(\alpha}h_{\rho)\lambda} + \frac{8\Lambda^{2}}{81}y^{2}h_{\alpha\rho} - \frac{2\Lambda^{2}}{81}y_{\alpha}y_{\rho}h \nonumber \\
&& + \frac{4\Lambda^{2}}{81}y^{\lambda}y_{(\alpha}h_{\rho)\lambda} - \frac{8\Lambda^{2}}{81}\eta_{\alpha\rho}y^{\lambda}y^{\gamma}h_{\lambda\gamma} \nonumber \\
&& - \frac{\Lambda}{9}y^{2}\Box h_{\alpha\rho} + \frac{\Lambda}{9}y^{\mu}y^{\nu}\partial_{\mu}\partial_{\nu}h_{\alpha\rho}
\end{eqnarray}
in which the convention for symmetrization is without factor of 1/2. Likewise, the right hand side of the equation is expanded in the normal neighborhood up to second order
\begin{eqnarray}\label{right}
&&(\Pi T)_{\alpha\rho}(y;x) \nonumber \\
&=& T_{\alpha\rho} + \frac{1}{3\mu^{2} - 2\Lambda}\Bigl\{\Bigl[\partial_{\rho}\partial_{\alpha} - \mu^{2}\eta_{\alpha\rho} + \frac{\Lambda}{3}\eta_{\alpha\rho}\Bigr]T \nonumber \\
&& + \Bigl[\frac{\Lambda}{3}\delta^{\lambda}_{\rho}y_{\alpha} + \frac{\Lambda}{9}\delta^{\lambda}_{\alpha}y_{\rho} - \frac{2\Lambda}{9}\eta_{\alpha\rho}y^{\lambda}\Bigr]\partial_{\lambda}T \nonumber \\
&& + \Bigl[\Bigl(\frac{\mu^{2}\Lambda}{9} - \frac{\Lambda^{2}}{27}\Bigr)\eta_{\alpha\rho}y^{2} - \Bigl(\frac{\mu^{2}\Lambda}{9} - \frac{\Lambda^{2}}{27}\Bigr)y_{\alpha}y_{\rho}\Bigr]T\Bigr\} \nonumber \\
&\equiv & (\Pi T_{0})_{\alpha\rho} + (\Pi T_{1})_{\alpha\rho} + (\Pi T_{2})_{\alpha\rho}
\end{eqnarray}
where the last line is defined in such a way that the subscript denotes the number of $y$'s (the adiabatic order). Thus, the solution to the equations of motion, evaluated up to second adiabatic order, is schematically written as
\begin{eqnarray}
h_{\alpha\rho}(x') &=& \int d^{4}x\sqrt{-\bar{g}}(G)_{\alpha\rho}^{\tau\sigma}(x',x)(\Pi T)_{\tau\sigma}(x) \nonumber \\
&\approx & \int d^{4}x\sqrt{-\bar{g}}\Bigl\{\sum_{i=0}^{2}(G_{i})_{\alpha\rho}^{\tau\sigma}(\Pi T_{0})_{\tau\sigma} \nonumber \\
&& + \sum_{i=0}^{1}(G_{i})_{\alpha\rho}^{\tau\sigma}(\Pi T_{1})_{\tau\sigma} + (G_{0})_{\alpha\rho}^{\tau\sigma}(\Pi T_{2})_{\tau\sigma}\Bigr\}
\end{eqnarray}
and the $G(x',x)$ is the adiabatic Green's function (subscript denoting the order) otbained by solving eq.(\ref{left}). Let us perform a Wick rotation $x^{0} \rightarrow ix^{0}$, for it is more convinent to compute the VPA in Euclidean space. The Green's function at zeroth adiabatic order is 
\begin{eqnarray}\label{zeroth order}
(G_{0})_{\alpha\rho}^{\tau\sigma}(x;x') &=& \int\frac{d^{4}k}{(2\pi)^{4/2}}\frac{\delta^{\tau\sigma}_{\alpha\rho}}{k^{2} + \mu^{2}}e^{ik\cdot(x-x')}\nonumber \\
\label{delta}
\text{where}\quad\delta^{\tau\sigma}_{\alpha\rho}&\equiv & \frac{1}{2}(\delta^{\tau}_{\alpha}\delta^{\sigma}_{\rho} - \delta^{\sigma}_{\alpha}\delta^{\tau}_{\rho})
\end{eqnarray}
which is the Feynman propagator $\Delta_{F} = 1/(k^{2} + \mu^{2})$ carrying indices after imposing constraints ($\nabla^{\alpha}h_{\alpha\rho} = \nabla_{\rho}h$ as well as taking the trace out) on the equations of motion in flat spacetime; note that this expression is valid only if the effective mass squared is positive, $\mu^{2}>0$. The first adiabatic order Green's functions are derived using the eq.(\ref{zeroth order}) and its defining equations are given as
\begin{eqnarray}\label{first order eqn}
&& [(-\Box + \mu^{2})\delta^{\lambda}_{\alpha}\delta^{\gamma}_{\rho}](G_{1})^{\tau\sigma}_{\lambda\gamma}(y) \nonumber \\
&=& -\frac{2\Lambda}{9}\delta^{\lambda}_{\alpha}\delta^{\gamma}_{\rho}y^{\beta}\partial_{\beta}(G_{0})^{\tau\sigma}_{\lambda\gamma}(y)+ \frac{2\Lambda}{9}y_{(\alpha}\delta^{\gamma}_{\rho)}\partial^{\lambda}(G_{0})^{\tau\sigma}_{\lambda\gamma}(y) \nonumber \\
&& - \frac{4\Lambda}{9}y^{\lambda}\delta^{\gamma}_{(\alpha}\partial_{\rho)}(G_{0})^{\tau\sigma}_{\lambda\gamma}(y)
\end{eqnarray}
whose solutions are given by 
\begin{eqnarray}\label{first order}
(G_{1})^{\tau\sigma}_{\alpha\rho}(k) &=& \frac{12\Lambda}{9}(\Delta_{F})(G_{0})^{\tau\sigma}_{\alpha\rho}(k) - \frac{4\Lambda}{9}k^{2}(\Delta_{F})^{2}(G_{0})^{\tau\sigma}_{\alpha\rho}(k) \nonumber \\
&& -\frac{2\Lambda}{9}K^{(\tau\sigma)}_{\alpha\rho}(\Delta_{F})^{3} - \frac{2\Lambda}{9}K^{(\tau\sigma)}_{\rho\alpha}(\Delta_{F})^{3}, \\
K^{(\tau\sigma)}_{\rho\alpha} &\equiv & \delta^{\tau}_{\rho}k^{\sigma}k_{\alpha} + \delta^{\sigma}_{\rho}k^{\tau}k_{\alpha}.
\end{eqnarray}
These are the first order curvature correction terms (which is linear in $\Lambda$) to the Green's functions with respect to the flat spacetime solution. The adiabatic second order Green's functions in momentum space, whose defining equations are deduced from the eq. (\ref{left}), can be derived in a similar fashion
\begin{widetext}
\begin{eqnarray}\label{second order}
(G_{2})^{\tau\sigma}_{\alpha\rho}(k) &=& \frac{64\Lambda^{2}}{81}(\Delta_{F})^{2}(G_{0})^{\tau\sigma}_{\alpha\rho}(k) + \frac{20\Lambda^{2}}{81}\eta^{\tau\sigma}\eta_{\alpha\rho}(\Delta_{F})^{3} - \frac{80\Lambda^{2}}{81}k^{2}(\Delta_{F})^{3}(G_{0})^{\tau\sigma}_{\alpha\rho}(k) + \frac{48\Lambda^{2}}{81}k^{4}(\Delta_{F})^{4}(G_{0})^{\tau\sigma}_{\alpha\rho}(k) \nonumber \\
&& - \frac{108\Lambda^{2}}{81}(K^{(\tau\sigma)}_{\alpha\rho} + K^{(\tau\sigma)}_{\rho\alpha})(\Delta_{F})^{4} - \frac{48\Lambda^{2}}{81}k^{\tau}k^{\sigma}\eta_{\alpha\rho}(\Delta_{F})^{4} - \frac{32\Lambda^{2}}{81}\eta^{\tau\sigma}k_{\alpha\rho}(\Delta_{F})^{4} \nonumber \\
&& + \frac{72\Lambda^{2}}{81}(K^{(\tau\sigma)}_{\alpha\rho} + K^{(\tau\sigma)}_{\rho\alpha})k^{2}(\Delta_{F})^{5} + \frac{96\Lambda^{2}}{81}k^{\tau}k^{\sigma}k_{\alpha}k_{\rho}(\Delta_{F})^{5}.
\end{eqnarray}
\end{widetext}
The detailed calculations can be found in the appendix. Therefore, the sum of eq.(\ref{zeroth order}), (\ref{first order}) and (\ref{second order}) is the inverse operator to the left hand side of the equations of motion, and it remains for us to put these pieces together into the VPA.

\section{Vacuum Persistence}
Unitarity of the linearized massive spin-2 theory can quantitatively be evaluated by using its partition function whose definition is given \cite{GibbonsHawking}
\begin{eqnarray}
\Braket{0|0}_{T} = \int(\mathcal{D}h_{\mu\nu})e^{iS[h;T]}
\end{eqnarray}
where $h_{\mu\nu}$ is the massive spin-2 field and its action $S[h;T]$ is given by eq.(\ref{action}), in which $T$ is taken to be a classical source; the partition function is to be bounded from above by unity if the model is healthy. Since it is easier to evalute this quantity in Euclidean spacetime, we will take the Wick rotation and see if the partition function in Euclidean spacetime is bounded in the same way. One might argue that we do not check unitarity as the computation will be done in Euclidean spacetime and we do not perform the analytic continuation back to real times. However, we can still use the result as an indication of the health of the model. The reasoning is simple: We compute a quantity at \textit{finite} euclidean time. If the Hamiltonian is bounded from below (a sign of health of the model), the difference between our result and a field theory at finite temperature should be small and our result should be smaller, but positive. Now, if the ground state is stable, at finite temperature the free energy $F=-k_B T \log(Z)$ (where $k_B$ is Boltzmann's constant, $T$ is the temperature, and $Z$ is the partition function) should be positive. This implies that $Z$ lies between $0$ and $1$. So, our result should be between $0$ and $1$ as well. \\

After taking the Wick rotation, the Jacobian factor $\sqrt{-\bar{g}}$ of the integration measure for the action is replaced by $\sqrt{\bar{g}_{E}}$ (where $E$ in the subscript denotes the Euclidiean signature), and the exponential part picks up another $i$ factor along with the action replaced by the Euclidean action
\begin{eqnarray}
\Braket{0|0}_{T} = \int(\mathcal{D}h_{\mu\nu})e^{-S_{E}[h;T]}.
\end{eqnarray}
This action can schematically be written as
\begin{eqnarray}
S_E = \int d^{4}x\sqrt{\bar{g}_{E}}\Bigl[-\frac{1}{2}h^{\mu\nu}\mathcal{K}_{\mu\nu, \rho\sigma}h^{\rho\sigma} + h_{\mu\nu}T^{\mu\nu}\Bigr]
\end{eqnarray}
where $\mathcal{K}_{\mu\nu, \rho\sigma}$ is pairwise symmetric ($\mu, \nu$ and $\rho, \sigma$) and contains all kinetic, mass and curvature operators. By changing the integration variable, $h_{\mu\nu} \rightarrow \chi_{\mu\nu} + (\mathcal{K}^{-1}T)_{\mu\nu}$ and after imposing the normalization condition such that the amplitude of the linear theory without any external source is nomalized to unity, the partition function is given below
\begin{eqnarray}
\Braket{0|0}_{T} &=& \exp\Bigl\{-\frac{1}{2}\int d^{4}x\sqrt{\bar{g}_{E}}\Bigl[T^{\mu\nu}(\mathcal{K}^{-1}T)_{\mu\nu}\Bigr]\Bigr\} \nonumber \\
&&\times \int(\mathcal{D}\chi_{\mu\nu})\exp\Bigl\{\frac{1}{2}\int d^{4}x\sqrt{\bar{g}_{E}}\Bigl[\chi^{\mu\nu}\mathcal{K}_{\mu\nu, \rho\sigma}\chi^{\rho\sigma}\Bigr] \nonumber \\
\Braket{0|0}_{T}^0 &\equiv & \exp\Bigl\{-\frac{1}{2}\int d^{4}x\sqrt{\bar{g}_{E}}\Bigl[T^{\mu\nu}(\mathcal{K}^{-1}T)_{\mu\nu}\Bigr]\Bigr\}
\end{eqnarray}
in which the second equality is the expression of the vacuum persistence amplitude (VPA); the superscript $0$ denotes the normalized partition function. Clearly, if a correct normalization is used, the overlap between any two states in quantum mechanics of a closed system should be bounded by one; otherwise, Born's rule is violated and the overlap cannot be interpreted in a probabilistic way. This holds irrespective of any manipulations performed with or within the system, so the expression in the above constrains the values of the exponent, provided the normalization is correct. This constraint is equivalent to imposing positivity of the Green's function or absence of tachyonic states in the spectrum. \\

This general form of VPA becomes more concrete as the equations of motion are solved using the power counting scheme $\Lambda y^{2}\ll1$. Inverse operator $\mathcal{K}^{-1}$ acting upon the external source in the integral is derived by solving the equation of motion, and the VPA is replaced by the following (the superscript notation of the VPA is dropped from now on for the context is clear)
\begin{eqnarray}
\Braket{0_{(2)}|0_{(2)}}_{T} &\approx & \exp\Bigl\{-\frac{1}{2}\int_{|x-x'|\ll\Lambda^{-1/2}} d^{4}x'\sqrt{\bar{g}'_{E}}d^{4}x\sqrt{\bar{g}_{E}} \nonumber \\
&&\times\Bigl[T^{\mu\nu}(x')G_{\mu\nu, \rho\sigma}(x',x)(\Pi T)^{\rho\sigma}(x)\Bigr]\Bigr\} 
\end{eqnarray}
where the integration is limited due to the approximation scheme discussed in the previous section and $G(x',x)$ is the adiabatic Green's function up to second order. Let us convert this expression to the Fourier space for it is easier to calculate the integral. The stress-energy tensor at $x$ within the normal neighborhood of $x'$ is
\begin{eqnarray}
T_{\mu\nu}(x) &=& T_{\mu\nu}(y;x) \nonumber \\
&=& \int\frac{d^{4}k}{(2\pi)^{4/2}\sqrt{\bar{g}_{E}}}\tilde{T}_{\mu\nu}(k)e^{ik\cdot x'}e^{ik\cdot y}
\end{eqnarray}
where $\tilde{T}_{\mu\nu}(k)$ is a Fourier amplitude of $T_{\mu\nu}(x)$. The VPA in Fourier space representation is, therefore, given order by order as follows
\begin{eqnarray}\label{VPA simple}
\Braket{0|0}_{T} &=&\exp\Bigl\{-\frac{1}{2}\int_{|k|\gg\Lambda^{1/2}}\frac{d^{4}k}{(2\pi)^{4}}\nonumber \\
&&\Bigl(\tilde{T}^{\alpha\rho}[G_{0} + G_{1} + G_{2}]^{\tau\sigma}_{\alpha\rho}(\Pi \tilde{T}_{0})_{\tau\sigma} \nonumber \\
&& + \tilde{T}^{\alpha\rho}[G_{0} + G_{1}]^{\tau\sigma}_{\alpha\rho}(\Pi \tilde{T}_{1})_{\tau\sigma}(k) \nonumber \\
&& + \tilde{T}^{\alpha\rho}(G_{0})^{\tau\sigma}_{\alpha\rho}(\Pi \tilde{T}_{2})_{\tau\sigma}(k)\Bigl)\Bigl\}
\end{eqnarray}
where the Jacobian factors, $\sqrt{\bar{g}_{E}}$ and $\sqrt{\bar{g}'_{E}}$, are canceled by the inverse Fourier transform of the sources at $x$ and $x'$. Since we are probing the neighborhood region around the origin $x'$ whose metric is Minkowskian, the continuity equation for the source is reduced to the flat one
\begin{eqnarray}
\nabla_{\alpha}T^{\alpha\rho} = 0 \rightarrow \partial_{\alpha}T^{\alpha\rho} =0
\end{eqnarray}
and thus $k_{\alpha}\tilde{T}^{\alpha\rho}(k) =0$ can be used to simplify the integrand of eq. (\ref{VPA simple}). The final form of the amplitude is given after plugging the eq.(\ref{zeroth order}), (\ref{first order}), (\ref{second order}) and (\ref{right}) into this expression and by specifying the static source as a point particle with its mass $\bar{M}$ localized on a spatial coordinate, (we treat the source as a classical object while the spin-2 field is treated quantum mechanically, which suffices at the linear level)
\begin{equation}\label{source}
T^{\alpha\rho}(x) \propto \bar{M}\delta^{\alpha}_{0}\delta^{\rho}_{0}\delta^{(3)}(\bold{x})
\end{equation}
which apparently obeys the continuity equation so that the VPA is given by (detailed calculations are found in the appendix)
\begin{widetext}
\begin{eqnarray}\label{VPA full}
\Braket{0|0}_{T} &=& \exp\Bigl\{-\frac{1}{2}\int_{|k|\gg\Lambda^{1/2}}\frac{d^{4}k}{(2\pi)^{4}}\Bigl[\frac{2\mu^{2} - 5\Lambda/3}{3\mu^{2} - 2\Lambda}(\Delta_{F}) + \frac{4\Lambda}{9(3\mu^{2} - 2\Lambda)}k^{2}(\Delta_{F})^{2} + \frac{\Lambda}{9(3\mu^{2} - 2\Lambda)}\Bigl(26\mu^{2} - \frac{62\Lambda}{3}\Bigr)(\Delta_{F})^{2} \nonumber \\
&& + \frac{32\Lambda^{2}}{81(3\mu^{2} - 2\Lambda)}k^{4}(\Delta_{F})^{4} + \frac{\Lambda}{9(3\mu^{2} - 2\Lambda)}\Bigl(\frac{104\Lambda}{9} - 16\mu^{2}\Bigr)k^{2}(\Delta_{F})^{3} \nonumber \\
&& + \Bigl[\frac{20\Lambda^{2}}{81} + \frac{\Lambda^{2}}{81(3\mu^{2} - 2\Lambda)}(48\mu^{2} - 60\Lambda)\Bigr](\Delta_{F})^{3} + \frac{\Lambda^{2}}{81(3\mu^{2} - 2\Lambda)}\Bigl(-112\mu^{2} + \frac{352\Lambda}{3}\Bigr)k^{2}(\Delta_{F})^{4} \nonumber \\
&& + \frac{48\Lambda^{2}(2\mu^{2} - 5\Lambda/3)}{81(3\mu^{2} - 2\Lambda)}k^{4}(\Delta_{F})^{5}\Bigl]|\tilde{T}(k)|^{2}\Bigr\}
\end{eqnarray}
\end{widetext}
expressed in the increasing order of the Feynman propagator. It is apparent that the first and second terms of the integrand are dominant in the high momentum regime as well as the curvature correction terms, which are higher orders in the Feynman propagator, do not dominate in the infrared regime, for the radius of the normal neighborhood is much smaller than the curvature radius; i.e., $k^{2}\gg\Lambda$. From eq. (\ref{source}), the Fourier amplitude of the external source is proportional to a delta function whose argument is a zeroth component of the four-momentum $T^{\mu\nu}(k)\sim \bar{M}\delta^{\mu}_{0}\delta^{\nu}_{0}\delta(k_{0})$, that the convergence of the VPA (i.e., the unitarity) boils down to the behaviour of the polynomial function in terms of the Feynman propagator. Combining the first two terms of the expression gives the leading order term of the polynomial function as
\begin{eqnarray}\label{leading order}
\frac{18\mu^{2} - 11\Lambda}{9(3\mu^{2} - 2\Lambda)}(\Delta_{F})^{1}
\end{eqnarray}
and by substituting $k^{2}$ with $x$ and $\mu^{2}$ with $M$, factoring out the coefficient of the leading order tem and taking the common denomimator by the fifth power of the Feynman propagator simplifies the expression inside the integral of eq. (\ref{VPA full}) as
\begin{equation}\label{integrand}
\frac{18M - 11\Lambda}{9(3M - 2\Lambda)}\cdot\frac{f(x)}{(x + M)^{5}}
\end{equation}
in which expression of the $f(x)$ is given by 
\begin{widetext}
\begin{eqnarray}\label{quartic function}
f(x) &=& (x + M)^{4} + \frac{\Lambda}{18M - 11\Lambda}\Bigl(22M - \frac{62\Lambda}{3}\Bigr)(x + M)^{3} + \frac{32\Lambda^{2}}{9(18M - 11\Lambda)}x^{2}(x + M) \nonumber \\
&& + \frac{\Lambda}{18M - 11\Lambda}\Bigl(\frac{104\Lambda}{9} - 16M\Bigr)x(x + M)^{2} + \frac{108\Lambda^{2}M - 100\Lambda^{3}}{9(18M - 11\Lambda)}(x + M)^{2} \nonumber \\
&& + \frac{\Lambda^{2}}{9(18M - 11\Lambda)}\Bigl(-112M + \frac{352\Lambda}{3}\Bigr)x(x + M) + \frac{48\Lambda^{2}(2M - 5\Lambda/3)}{9(18M - 11\Lambda)}x^{2}.
\end{eqnarray}
\end{widetext}
It is apparent that $(x+M)^{-5}$ in eq. (\ref{integrand}) is positive because $x>0$ and we impose $M\geq0$, so it is necessary for us to analyze the behaviour of the polynomial function $f(x)$ up to a quartic order in $x$. In order to show the positivity of $f(x)$ in its domain, one has to first find its extrema (minima, in particular). Extrema are evaluated by taking a first order differentiation with respect to $x$ and solving for its roots. Since there is only one real extremum of the function, it is the point where the minimum occurs (detailed analysis is found in the appendix), given that the coefficient of the leading order term, eq. (\ref{leading order}), is positive. It turns out that the point at which the minium occurs lies outside the domain of the function ($x>0$) that the actual minimum of the function is at the infrared (IR) cutoff of the theory. The IR cutoff is at the scale much greater than the curvature scale, and one can take, for instance, $3\Lambda$ (or much greater) as the lower bound of the integral, and it is shown that $f(3\Lambda)$ is strictly positive (see appendix). This implies that the integrand of the eq.(\ref{VPA simple}) is strictly positive in its domain, and the VPA, thereafter, converges to a finite value between zero and one.  \\

If we impose that the VPA does not violate unitarity to leading order, we thus obtain a healthy theory if and only if
\begin{eqnarray}
\mu^{2}>\frac{2\Lambda}{3}\quad &\text{and}&\quad\mu^{2}>\frac{11\Lambda}{18} \\
\text{or}\quad\mu^{2}<\frac{2\Lambda}{3}\quad &\text{and}&\quad 0<\mu^{2}<\frac{11\Lambda}{18}.
\end{eqnarray}
Rewritten in terms of the bare mass $m$ and the cosmological constant, these equations are
\begin{equation}\label{our bound}
m^{2} + 4\xi\Lambda > \frac{2\Lambda}{3}\quad\leftrightarrow\quad m^{2}>\Bigl(\frac{2}{3} - 4\xi\Bigr)\Lambda
\end{equation}
and 
\begin{equation}
0<m^{2} + 4\xi\Lambda <\frac{11\Lambda}{18}\quad\leftrightarrow\quad -4\xi\Lambda < m^{2}<\Bigl(\frac{11}{18} - 4\xi\Bigr)\Lambda.
\end{equation}
Depending on the values of $\xi$, certain cosmological constants would become unstable. However, there is a critical value $\xi_0 = \frac{1}{6}$, which is called the conformal coupling \cite{ParkerToms, Birell}, at which \emph{all} de Sitter and Minkowski spacetimes are stable, as then the first condition reduces to $m^{2}>0$ (which is the well-known bound from Minkowski). \\

One further comment is in order: Clearly, due to the special choice of the source, it is not unequivocal whether the result obtained here actually gives the correct bound. There might be a set of unstable states which simply do not couple to the source considered here; insofar this analysis clearly does not have the claim to completeness as Higuchi's considerations \cite{HiguchiBound}. However, setting the coupling strength to zero, we recover Higuchi's celebrated result, so at least a relevant subset of all ``dangerous'' unstable states couples to source considered here. Our bound, in view of this, should be read as a necessary, but not a sufficient, condition to obtain stability of spacetime under perturbations described by a massive spin-2 field. The restriction to a (possibly) small local neighbourhood only emphasizes this last point. 

\section{Conclusion and Discussion}
\subsection{Mass Bound}
In his paper \cite{HiguchiBound}, Higuchi showed the mass bound of the spin-2 theory by the canonical quantization instead of the path-integral formulation. A class of equal-time commutation relation gives rise to the lower bound of the mass squared, $m^{2}>2/3\Lambda$, when the negative norm state is avoided. However, this lower bound physically implies that the quanta of the massive spin-2 field are able to destabilize de Sitter spacetime, which possibly stems from the improper covariantization of the linearized theory; a particular choice of the coupling between the field and background spacetime, $\xi = 0$. There is no justification of chooing the minimal coupling when a general covariant theory in curved background is considered. One can recover eq. (\ref{our bound}) when the non-minimal coupling term is included in the action, and perform the equal-time commutation relation just as shown in the Higuchi's paper \cite{HiguchiBound}. The corrected mass bound implies the absence of the forbidden range of the mass when quantizing the theory. A notion of the global vacuum is defined on a spacelike hypersurface in order to derive the bound, whereas we used the adiabatic Green's function and adiabatic vacuum up to second order of the manifold in order to compute the VPA. So it is legitimate if one raise a doubt if the approach in this papar is fully valid. Whether the physical set-up is global or local, however, as far as any valid vacuum to vacuum transition is concerned, eq. (\ref{our bound}) is derived by imposing the unitarity at the local region, and the analysis here is consistent at the linear level. Had the theory already broken down in a local region, there would be no hope for the same theory to recover globally.\\

\subsection{Anti-de Sitter Spacetime}
One can ask whether the anti-de Sitter spacetime is also a proper solution the massive spin-2 theory, given the de Sitter is its solution, for the previous analysis was done in the maximally symmetric 4-dimension spacetime regardless of the sign of the curvature. The VPA in the anti-de Sitter spacetime is altered simply by replacing $\Lambda$ with $-|\Lambda|$ in eq. (\ref{VPA simple}) as following
\begin{widetext}
\begin{eqnarray}
&& \braket{0_{(2)}|0_{(2)}}_{T} \nonumber \\ 
&=& \exp\Bigl\{-\frac{1}{2}\int_{k^{2}\gg |\Lambda|}\frac{d^{4}k}{(2\pi)^{4}}\Bigl[\frac{2\mu^{2} + 5|\Lambda|/3}{3\mu^{2} + 2|\Lambda|}(\Delta_{F}) - \frac{4|\Lambda|}{9(3\mu^{2} + 2|\Lambda|)}k^{2}(\Delta_{F})^{2} \nonumber \\
&& - \frac{|\Lambda|}{9(3\mu^{2} + 2|\Lambda|)}\Bigl(26\mu^{2} + \frac{62|\Lambda|}{3}\Bigr)(\Delta_{F})^{2} + \frac{32|\Lambda|^{2}}{81(3\mu^{2} + 2|\Lambda|)}k^{4}(\Delta_{F})^{4} - \frac{|\Lambda|}{9(3\mu^{2} + 2|\Lambda|)}\Bigl(-\frac{104|\Lambda|}{9} - 16\mu^{2}\Bigr)k^{2}(\Delta_{F})^{3} \nonumber \\
&& + \Bigl[\frac{20|\Lambda|^{2}}{81} + \frac{|\Lambda|^{2}}{81(3\mu^{2} + 2\Lambda)}(48\mu^{2} + 60\Lambda)\Bigr](\Delta_{F})^{3} + \frac{|\Lambda|^{2}}{81(3\mu^{2} + 2|\Lambda|)}\Bigl(-112\mu^{2} - \frac{352|\Lambda|}{3}\Bigr)k^{2}(\Delta_{F})^{4} \nonumber \\
&& + \frac{48|\Lambda|^{2}(2\mu^{2} + 5|\Lambda|/3)}{81(3\mu^{2} + 2|\Lambda|)}k^{4}(\Delta_{F})^{5}\Bigl]|\tilde{T}(k)|^{2}\Bigr\}
\end{eqnarray}
\end{widetext}
and the expression of the effective mass is altered by $\mu^{2} = m^{2} - 4\xi|\Lambda|$, which implies that the $\mu^{2}$ is no longer treated as a non-negative parameter. However, the Feynman parameter does not suffer from hitting poles due to the domain of the integrand, $k^{2} \gg |\Lambda|$, so that we can perform the same unitarity analysis on this amplitude as in the previous section. The mass bound of the spin-2 field comes from the positivity of the coefficient of the leading order term
\begin{eqnarray}\label{ads bound}
\frac{18\mu^{2} + 11|\Lambda|}{9(3\mu^{2} + 2|\Lambda|)}&>&0\quad\rightarrow\quad \mu^{2} > -\frac{11|\Lambda|}{18} \nonumber \\
\quad\leftrightarrow\quad m^{2}&>&\Bigl(4\xi - \frac{11}{18}\Bigr)|\Lambda|
\end{eqnarray}
and the lower bound to the mass is ruled out by the similar reasoning given the previous section. This mass bound looks again non-trivially related to the background curvature when the coupling strength is set to equal to one. We can constrain this value within a range $-1\leq\xi<11/72\sim 1/7$, and the mass squared is again bounded from below by a negative quantity. Notice that the massive spin-2 field at the linear level is weakly coupled to the background geometry and that, if the coupling strength is adjusted in the de Sitter spacetime, the same theory is to be held in the anti-de Sitter spacetime with the similar coupling constant, given that both are proper solutions to the massive gravity. This compatibility between the two maximally symmetric spacetimes suggests the restriction on the value of the coupling between 1/6 and 1/7 in such a way that the constant becomes slightly greater than (or equal to) 1/6 when the background has a positive curvature and it is slightly smaller than 1/7 if the background has a negative curvature. Imagine a free massive graviton is propagating in the de Sitter background, and all of a sudden, the background is locally changed into the anti-de Sitter spacetime due to the presence of a source, such as a black hole \cite{BH}. The same graviton is also to be freely propagating in this background regardless of its curvature, and these two bounds of the coupling constant ensure the safety of the graviton in the maximally symmetric spacetimes. Had the mass had a certain forbidden region, as is claimed in \cite{HiguchiBound}, the quantum description of the graviton with a fixed value of the mass is to be broken down by the unitarity violation, when it enters another region of the constant curvature whose magnitude is greater than the graviton mass. This violation implies that either the background is unstable, the degrees of freedom are altered by the change of the background, inclusion of the higher order terms might restore the unitarity, or the unitarity can be restored in some other ways. In any case, the effective description is to be modified in some way or the other when the unitarity violation occurs. 

\subsection{FRW Universe}
The unitarity analysis given above is slightly modified when the background is generalized from maximal symmetry in spacetime merely to space, such as a homogeneous and isotropic universe (Friedmann-Robertson-Walker or FRW universe in short). Recall the relation between the normal coordinates (or the wavelength of graviton) $y$ and the Hubble time $H$, for a time parameter in FRW universe is $y^{2}\ll H^{-2}$. The Hubble parameter is a constant value in de Sitter and anti-de Sitter spacetimes, but it becomes a time-dependent parameter in the comoving frame once the spacetime symmetry is abandoned; i.e., $H^{2}(t)y^{2}\ll 1$ is the adiabatic parameter. When the universe is assumed to be spatially flat, which is what we are observing today \cite{Liddle, Stompor, Spergel}, and its expansion rate is slow ($\dot{H}\ll H^{2}$), the adiabatic parameter can also be written in terms of the curvature scalar
\begin{eqnarray}
\mathcal{R}(t)y^{2}\ll 1
\end{eqnarray}
which clearly differs from de Sitter (or anti-de Sitter) case such that the Ricci scalar is now time dependent. Slow expansion ($\dot{H}\ll H^{2}$) of the universe implies that the significant expansion takes place in the time span much larger than the Hubble time), so that we can mimic the similar analysis in the maximally symmetric spacetimes, for the curvature scalar is almost a constant in the normal neighborhood region. Therefore, the mass bound in FRW universe naturally arises when unitarity is imposed
\begin{eqnarray}
m^{2}>\Bigl(\frac{2}{3} - 4\xi\Bigr)H^{2}(t)
\end{eqnarray}
and this agrees with the literature \cite{HiguchiBound, Berkhahn}, when the coupling strength is minimal. This bound seemingly runs due to the time dependence of the Hubble parameter. We can treat the Hubble parameter as a constant in this context if the local region of the spacetime is restricted enough, and conclude that the mass bound is again absent in the slowly expanding and spatially flat FRW universe. 
  
\section{Appendix}
\subsection{Adiabatic Green's function}
As taking the Fourier tansform of the eq. (\ref{first order eqn}), the first term on the right hand side of the equation is expressed as following
\begin{eqnarray}
\frac{2\Lambda}{9}\tilde{\partial}^{\beta}\Bigl[k_{\beta}\frac{\delta^{\tau\sigma}_{\alpha\rho}}{k^{2} + \mu^{2}}\Bigr] &=& \frac{8\Lambda}{9}\cdot\frac{\delta^{\tau\sigma}_{\alpha\rho}}{k^{2} + \mu^{2}} - \frac{4\Lambda}{9}\cdot\frac{k^{2}\delta^{\tau\sigma}_{\alpha\rho}}{(k^{2} + \mu^{2})^{2}} \nonumber \\
\text{where}\quad \tilde{\partial}^{\beta} &=& \frac{\partial}{\partial k_{\beta}}
\end{eqnarray}
and $\delta^{\tau\sigma}_{\alpha\rho}$ is defined in the eq. (\ref{delta}). The second and the third terms are transformed in a similar fashion
\begin{eqnarray}
-\frac{2\Lambda}{9}\tilde{\partial}_{(\alpha}\Bigl[k^{\lambda}\frac{\delta^{\tau\sigma}_{\lambda|\rho)}}{k^{2} + \mu^{2}}\Bigr] &=& -\frac{4\Lambda}{9}\Bigl[\frac{\delta^{\tau\sigma}_{\alpha\rho}}{k^{2} + \mu^{2}} - \frac{K^{(\tau\sigma)}_{\alpha\rho}}{(k^{2} + \mu^{2})^{2}}\Bigr] \nonumber \\
\frac{4\Lambda}{9}\tilde{\partial}^{\lambda}\Bigl[\frac{\delta^{\tau\sigma}_{\lambda(\alpha}k_{\rho)}}{k^{2} + \mu^{2}}\Bigr] &=& \frac{8\Lambda}{9}\Bigl[\frac{\delta^{\tau\sigma}_{\alpha\rho}}{k^{2} + \mu^{2}} - \frac{K^{(\tau\sigma)}_{\alpha\rho}}{(k^{2} + \mu^{2})^{2}}
\end{eqnarray}
in which the definition of $K^{(\tau\sigma)}_{\rho\alpha}$ is given in eq. (\ref{first order}), and thus, the first order Green's functions are derived as in the eq. (\ref{first order}). After solving for the first order Green's function, the defining equations of the second order Green's functions are given as 
\begin{widetext}
\begin{eqnarray}\label{first term G}
(-\Box + \mu^{2})\delta^{\lambda}_{\alpha}\delta^{\gamma}_{\rho}(G_{2})^{\tau\sigma}_{\lambda\gamma}(y) &=& -\frac{2\Lambda}{9}\delta^{\lambda}_{\alpha}\delta^{\gamma}_{\rho}y^{\beta}\partial_{\beta}(G_{1})^{\tau\sigma}_{\lambda\gamma}(y) + \frac{2\Lambda}{9}y_{(\alpha}\delta^{\gamma}_{\rho)}\partial^{\lambda}(G_{1})^{\tau\sigma}_{\lambda\gamma}(y) - \frac{4\Lambda}{9}y^{\lambda}\delta^{\gamma}_{(\alpha}\partial_{\rho)}(G_{1})^{\tau\sigma}_{\lambda\gamma}(y) \nonumber \\
&& - \frac{8\Lambda^{2}}{81}y^{2}\delta^{\lambda}_{\alpha}\delta^{\gamma}_{\rho}(G_{0})^{\tau\sigma}_{\lambda\gamma}(y) + \frac{2\Lambda^{2}}{81}y_{\alpha}y_{\rho}\eta^{\lambda\gamma}(G_{0})^{\tau\sigma}_{\lambda\gamma}(y) - \frac{4\Lambda^{2}}{81}y^{\lambda}y_{(\alpha}\delta^{\gamma}_{\rho)}(G_{0})^{\tau\sigma}_{\lambda\gamma}(y) \nonumber \\
&& + \frac{8\Lambda^{2}}{81}\eta_{\alpha\rho}y^{\lambda}y^{\gamma}(G_{0})^{\tau\sigma}_{\lambda\gamma}(y) + \frac{2\Lambda}{9}y^{2}\Box\delta^{\tau}_{\alpha}\delta^{\sigma}_{\rho}(G_{0})^{\tau\sigma}_{\lambda\gamma}(y) - \frac{\Lambda}{9}y^{\mu}y^{\nu}\partial_{\mu}\partial_{\nu}\delta^{\tau}_{\alpha}\delta^{\sigma}_{\rho}(G_{0})^{\tau\sigma}_{\lambda\gamma}(y).
\end{eqnarray}
Notice that the last two terms on the right hand side of the equations cancel each other because the zeroth order Green's functions are Lorentz invariant; i.e., they are proportional to $\delta^{\tau\sigma}_{\lambda\gamma}y^{2}$ that both terms are identical with the opposite sign when we plug this expression into the equations. The first term on the right hand side of the equation in Fourier space in terms of the zeroth order Green's functions is given by (let us keep 1/81 as a common numerical factor in front of each term from now on)
\begin{eqnarray}\label{first term G2}
\frac{2\Lambda}{9}\tilde{\partial}^{\beta}[k_{\beta}(G_{1})^{\tau\sigma}_{\lambda\gamma})](k) &=& \frac{96\Lambda^{2}}{81}(\Delta_{F})(G_{0})^{\tau\sigma}_{\alpha\rho}(k) - \frac{144\Lambda^{2}}{81}k^{2}(\Delta_{F})^{2}(G_{0})^{\tau\sigma}_{\alpha\rho}(k) + \frac{48\Lambda^{2}}{81}k^{4}(\Delta_{F})^{3}(G_{0})^{\tau\sigma}_{\alpha\rho}(k) \nonumber \\
&& - \frac{24\Lambda^{2}}{81}(K^{(\tau\sigma)}_{\alpha\rho} + K^{(\tau\sigma)}_{\rho\alpha})(\Delta_{F})^{3} + \frac{24\Lambda^{2}}{81}(K^{(\tau\sigma)}_{\alpha\rho} + K^{(\tau\sigma)}_{\rho\alpha})k^{2}(\Delta_{F})^{4}
\end{eqnarray}
and likewise the second and third terms are transformed as
\begin{eqnarray}\label{second term G2}
-\frac{2\Lambda}{9}\tilde{\partial}_{(\alpha}\delta^{\gamma}_{\rho)}[k^{\lambda}(G_{1})^{\tau\sigma}_{\lambda\gamma}(k)] &=& -\frac{48\Lambda^{2}}{81}(\Delta_{F})(G_{0})^{\tau\sigma}_{\alpha\rho}(k) + \frac{32\Lambda^{2}}{81}k^{2}(\Delta_{F})^{2}(G_{0})^{\tau\sigma}_{\alpha\rho}(k) + \frac{72\Lambda^{2}}{81}(K^{(\tau\sigma)}_{\alpha\rho} + K^{(\tau\sigma)}_{\rho\alpha})(\Delta_{F})^{3} \nonumber \\
&& + \frac{16\Lambda^{2}}{81}k^{\tau}k^{\sigma}\eta_{\alpha\rho}(\Delta_{F})^{3} - \frac{48\Lambda^{2}}{81}(K^{(\tau\sigma)}_{\alpha\rho} + K^{(\tau\sigma)}_{\rho\alpha})k^{2}(\Delta_{F})^{4} - \frac{96\Lambda^{2}}{81}k^{\tau}k^{\sigma}k_{\alpha}k_{\rho}(\Delta_{F})^{4} \\
\label{third term G2}
\frac{4\Lambda}{9}\tilde{\partial}^{\lambda}\delta^{\gamma}_{(\alpha}[k_{\rho)}(G_{1})^{\tau\sigma}_{\lambda\gamma}(k)] &=& \frac{96\Lambda^{2}}{81}(\Delta_{F})(G_{0})^{\tau\sigma}_{\alpha\rho}(k) - \frac{32\Lambda^{2}}{81k^{2}}(\Delta_{F})^{2}(G_{0})^{\tau\sigma}_{\alpha\rho}(k) - \frac{172\Lambda^{2}}{81}(K^{(\tau\sigma)}_{\alpha\rho} + K^{(\tau\sigma)}_{\rho\alpha})(\Delta_{F})^{3} \nonumber \\
&& - \frac{16\Lambda^{2}}{81}\eta^{\tau\sigma}k_{\alpha}k_{\rho}(\Delta_{F})^{3} + \frac{96\Lambda^{2}}{81}(K^{(\tau\sigma)}_{\alpha\rho} + K^{(\tau\sigma)}_{\rho\alpha})k^{2}(\Delta_{F})^{4} + \frac{192\Lambda^{2}}{81}k^{\tau}k^{\sigma}k_{\alpha}k_{\rho}(\Delta_{F})^{4}.
\end{eqnarray}
At last, the remaining terms (from the fourth to the seventh) are given by
\begin{eqnarray}\label{fourth term G2}
\frac{8\Lambda^{2}}{81}\tilde{\partial}^{\lambda}\tilde{\partial}_{\lambda}\Bigl(\frac{\delta^{\tau\sigma}_{\alpha\rho}}{k^{2} + \mu^{2}}\Bigr) &=& -\frac{64\Lambda^{2}}{81}(\Delta_{F})(G_{0})^{\tau\sigma}_{\alpha\rho}(k) + \frac{64\Lambda^{2}}{81}k^{2}(\Delta_{F})^{2}(G_{0})^{\tau\sigma}_{\alpha\rho}(k)\\
\label{fifth term G2}
-\frac{2\Lambda^{2}}{81}\tilde{\partial}_{\alpha}\tilde{\partial}_{\rho}\Bigl(\frac{\eta^{\lambda\gamma}\delta^{\tau\sigma}_{\lambda\gamma}}{k^{2} + \mu^{2}}\Bigr) &=& \frac{4\Lambda^{2}}{81}\eta^{\tau\sigma}\eta_{\alpha\rho}(\Delta_{F})^{2} - \frac{16\Lambda^{2}}{81}\eta^{\tau\sigma}k_{\alpha}k_{\rho}(\Delta_{F})^{3} \\
\label{sixth term G2}
\frac{4\Lambda^{2}}{81}\tilde{\partial}^{\lambda}\tilde{\partial}_{(\alpha}\Bigl(\delta^{\gamma}_{\rho)}\frac{\delta^{\tau\sigma}_{\lambda\gamma}}{k^{2} + \mu^{2}}\Bigr) &=& -\frac{16\Lambda^{2}}{18}(\Delta_{F})(G_{0})^{\tau\sigma}_{\alpha\rho}(k) + \frac{16\Lambda^{2}}{81}(K^{(\tau\sigma)}_{\alpha\rho} + K^{(\tau\sigma)}_{\rho\alpha})(\Delta_{F})^{3} \\
\label{seventh term G2}
-\frac{8\Lambda^{2}}{81}\eta_{\alpha\rho}\tilde{\partial}^{\lambda}\tilde{\partial}^{\gamma}\Bigl(\frac{\delta^{\tau\sigma}_{\lambda\gamma}}{k^{2} + \mu^{2}}\Bigr) &=& \frac{16\Lambda^{2}}{81}\eta_{\alpha\rho}\eta^{\tau\sigma}(\Delta_{F})^{2} - \frac{64\Lambda^{2}}{81}\eta_{\alpha\rho}k^{\tau}k^{\sigma}(\Delta_{F})^{3}
\end{eqnarray}
from which one can solve for the second order Green's function by summing eq. (\ref{first term G2}), (\ref{second term G2}), (\ref{third term G2}), (\ref{fourth term G2}), (\ref{fifth term G2}), (\ref{sixth term G2}), and (\ref{seventh term G2}) and divide it by the Feynman propagator, whose expression is given in eq. (\ref{second order}). 
\subsection{VPA expansion}
Let us calculate the term by term in the integral from the eq. (\ref{VPA simple}) with the given Green's function (eq. (\ref{zeroth order}), eq. (\ref{first order}) and eq. (\ref{second order})). Referring to the eq. (\ref{right}), the last term from the eq. (\ref{VPA simple}) in the Fourier space is given by
\begin{eqnarray}
\tilde{T}^{\alpha\rho}(k)(G_{0})^{\tau\sigma}_{\alpha\rho}(\Pi\tilde{T}_{2})_{\tau\sigma}(-k) &=& \frac{3\mu^{2}\Lambda - \Lambda^{2}}{27(3\mu^{2} - 2\Lambda)}\tilde{T}^{\alpha\rho}[2\eta_{\alpha\rho}(\Delta_{F})^{2} - 8\eta_{\alpha\rho}k^{2}(\Delta_{F})^{2} + 8k_{\alpha}k_{\rho}(\Delta_{F})^{3}]\tilde{T}(-k) \nonumber \\
&=& \Bigl[\Bigl(\frac{6\mu^{2}\Lambda - 2\Lambda^{2}}{27(3\mu^{2} - 2\Lambda)}\Bigr)(\Delta_{F})^{2} - \Bigl(\frac{24\mu^{2}\Lambda - 8\Lambda^{2}}{27(3\mu^{2} - 2\Lambda)}\Bigr)k^{2}(\Delta_{F})^{3}\Bigr]|\tilde{T}(k)|^{2}
\end{eqnarray}
in which it is clear that the conservation of the source $k_{\alpha}\tilde{T}^{\alpha\rho}(k) = 0$ is used on the second line. One can perform the similar calculation on the second term in the eq. (\ref{VPA simple}) as follows
\begin{eqnarray}
\tilde{T}^{\alpha\rho}(k)[(G_{0}) + (G_{1})]^{\tau\sigma}_{\alpha\rho}(\Pi\tilde{T}_{1})_{\tau\sigma}(-k) &=& \frac{1}{3\mu^{2} - 2\Lambda}\Bigl[\frac{4\Lambda}{9}k^{2}(\Delta_{F})^{2} + \frac{40\Lambda^{2}}{81}k^{2}(\Delta_{F})^{3} - \frac{16\Lambda^{2}}{81}k^{4}(\Delta_{F})^{4}\Bigr]|\tilde{T}(k)|^{2}.
\end{eqnarray}
The first term in the eq. (\ref{VPA simple}) is given order by order such that the zeroth order term is given by
\begin{eqnarray}
\tilde{T}^{\alpha\rho}(k)(G_{0})^{\tau\sigma}_{\alpha\rho}(\Pi\tilde{T}_{0})_{\tau\sigma}(-k) &=& \Bigl[|\tilde{T}^{\alpha\rho}|^{2} - \frac{1}{3\mu^{2} - 2\Lambda}\Bigl(\mu^{2} - \frac{\Lambda}{3}\Bigr)|\tilde{T}|^{2}\Bigr](\Delta_{F}) = \frac{2\mu^{2} - 5\Lambda/3}{3\mu^{2} - 2\Lambda}(\Delta_{F})|\tilde{T}(k)|^{2}
\end{eqnarray}
and the expression of the localized source (eq. (\ref{source}) was used for the second equality. Likewise, the first order Green's function contribution to it is 
\begin{eqnarray}
\tilde{T}^{\alpha\rho}(k)(G_{1})^{\tau\sigma}_{\alpha\rho}(\Pi\tilde{T}_{0})_{\tau\sigma}(-k) &=& \Bigl[\frac{12\Lambda(2\mu^{2} - 5\Lambda/3)}{9(3\mu^{2} - 2\Lambda)}(\Delta_{F})^{2} - \frac{4\Lambda(2\mu^{2} - 5\Lambda/3)}{9(3\mu^{2} - 2\Lambda)}k^{2}(\Delta_{F})^{3}\Bigr]|\tilde{T}(k)|^{2}
\end{eqnarray}
where the identity $K^{(\tau\sigma)}_{\alpha\rho}\tilde{T}^{\alpha\rho}(k) = 0$ from the conservation equation is used. The second order contribution is the following
\begin{eqnarray}
\tilde{T}^{\alpha\rho}(k)(G_{2})^{\tau\sigma}_{\alpha\rho}(\Pi\tilde{T}_{0})_{\tau\sigma}(-k) &=& \Bigl\{\Bigl[\frac{20\Lambda^{2}}{81} + \frac{\Lambda^{2}(48\mu^{2} - 60\Lambda)}{81(3\mu^{2} - 2\Lambda)} \Bigr](\Delta_{F})^{3} + \frac{\Lambda^{2}}{81(3\mu^{2} - 2\Lambda)}\Bigl(-k^{2}(\Delta_{F})^{3} + k^{4}(\Delta_{F})^{4}\Bigr) \nonumber \\
&& + \frac{\Lambda^{2}}{81(3\mu^{2} - 2\Lambda)}\Bigl(-112\mu^{2} + \frac{352\Lambda}{3}\Bigr)k^{2}(\Delta_{F})^{4}  + \frac{48\Lambda^{2}(2\mu^{2} - 5\Lambda/3)}{81(3\mu^{2} - 2\Lambda)}k^{4}(\Delta_{F})^{5}\Bigr\}|\tilde{T}|^{2}
\end{eqnarray}
so that the first term of the eq. (\ref{VPA simple}) is altogether given by
\begin{eqnarray}
\tilde{T}^{\alpha\rho}(k)(G_{0} + G_{1} + G_{2})^{\tau\sigma}_{\alpha\rho}(\Pi T_{0})_{\tau\sigma}(-k) &=& \Bigl\{\frac{2\mu^{2} - 5\Lambda/3}{3\mu^{2} - 2\Lambda}(\Delta_{F}) + \frac{12\Lambda(2\mu^{2} - 5\Lambda/3}{9(3\mu^{2} - 2\Lambda)}(\Delta_{F})^{2} - \frac{4\Lambda(2\mu^{2} - 10\Lambda/9)}{9(3\mu^{2} - 2\Lambda)}k^{2}(\Delta_{F})^{3} \nonumber \\ 
&& + \Bigl[\frac{20\Lambda^{2}}{81} + \frac{\Lambda^{2}(48\mu^{2} - 60\Lambda)}{81(3\mu^{2} - 2\Lambda)}\Bigr](\Delta_{F})^{3} + \frac{\Lambda^{2}(-112\mu^{2} + 352\Lambda/3)}{81(3\mu^{2} - 2\Lambda)}k^{2}(\Delta_{F})^{4} \nonumber \\ 
&& + \frac{48\Lambda^{2}}{81(3\mu^{2} - 2\Lambda)}k^{4}(\Delta_{F})^{4} + \frac{48\Lambda^{2}(2\mu^{2} - 5\Lambda/3)}{81(3\mu^{2} - 2\Lambda)}k^{4}(\Delta_{F})^{5} \Bigr\}|\tilde{T}|^{2}
\end{eqnarray}
\end{widetext}
and the VPA in the momentum space is written in the eq. (\ref{VPA full}).

\subsection{Bound Analysis}
\begin{figure}[h]
\includegraphics[width=3.5cm]{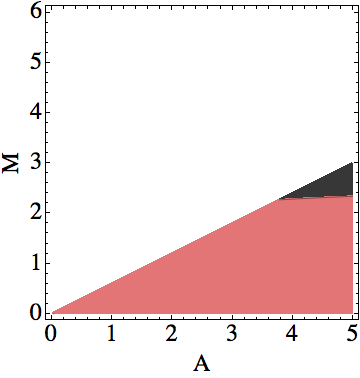}
\caption{$\Lambda (=A)$ versus $M$ when $M<11\Lambda/18$. Colored regions are the minima $x_0$, and the contour line corresponds to $x_0 = -2$. The rest are also negative values, which means the minima is out of domain of the function $f(x)$.}
\label{x when M<11/18}
\end{figure}

As the quartic function $f(x)$ is expressed in eq. (\ref{quartic function}) and its positivity on its domain are concerned, its extrema are found by taking first order differentiation with respect to $x$ and setting it equal to zero (can be checked by Mathematica). Two extrema are complex values, and there only remains only one real extremum, $x_{0}$, which is the point at which the minimum occurs. Since the condition
\begin{equation}
\frac{18M - 11\Lambda}{9(3M - 2\Lambda)}>0
\end{equation}
is assumed from the beginning, two different plots of the minima $x_{0}$ with respect to the two parameters, $\Lambda$ and $M$, are to be analyzed: 
\begin{figure}[h]
\includegraphics[width=3.5cm]{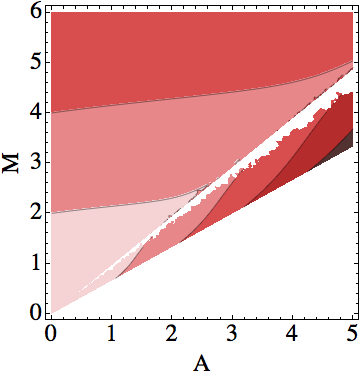}
\caption{$\Lambda (=A)$ versus $M$ when $M>2\Lambda/3$. Colored regions signify the minima $x_0$, with the contour lines progress as -2, -4, and so forth; i.e., the evaluated minima are out of the domain of the function $f(x)$.}
\label{x when M>2/3}
\end{figure}
The contour map for $M>2\Lambda/3$ is given in FIG. \ref{x when M>2/3}, in which the horizontal axis $A$ is the value of the curvature constant normalized by the corresponding unit and vertical axis is the value of $M$ normalized likewise. There is a forbidden region between the colored ones, and this is where $x_{0}$ is not well-defined from the condition $f'(x) = 0$. Contour line at the bottom corresponds to $-2$, the next line to $-4$ and so forth, from which we conclude $x_{0}<0$. The second plot for the other condition, $M<11\Lambda/18$, has a simpler feature where its contour line is at $-2$ and the rest are also negative values (see FIG. \ref{x when M<11/18}). 
These two contour maps inform us that the minimum of $f(x)$ occurs when $x_{0}<0$, which is outside the domain of the function, so it is necessary for us to check if the function is positive in its domain when the two conditions of the parameters are assumed. \\

\begin{figure}[h]
\includegraphics[width=3.5cm]{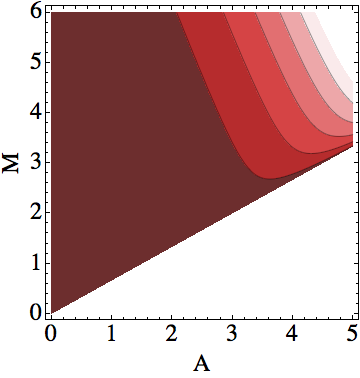}
\caption{$f(3\Lambda)$ when $M>2\Lambda/3$. Colored regions are the values of $f(3\Lambda)$, and the contour lines progress as 25000, 50000, and so forth, which means $f(x)$ is strictly positive in its domain, given $M>2\Lambda/3$.}
\label{f when M>2/3}
\end{figure}
One can choose an arbitrary value of $x$ as long as it is much greater than the curvature constant $\Lambda$ (recall $x\gg\Lambda$ from the adiabatic expansion), so let us choose $3\Lambda$ as a test. A contour plot for $f(3\Lambda)$ in terms of $\Lambda$ and $M$ for $M>2\Lambda/3$ is given in FIG. \ref{f when M>2/3} such that the first contour line corresponds to $f(3\Lambda) = 25000$, the second line to 50,000, and so forth; $f(x)$ is strictly positive for $x=3\Lambda$.  \\

\begin{figure}[h]
\includegraphics[width=3.5cm]{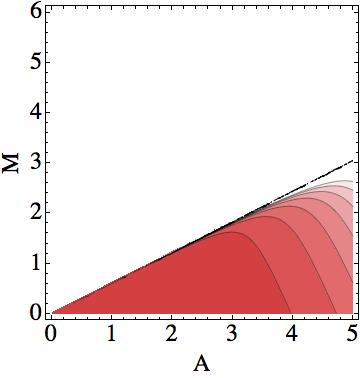}
\caption{$f(3\Lambda)$ when $M<11\Lambda/18$. As in FIG. 3., colored regions correspond to the values of $f(3\Lambda)$, and the contour lines are given as 25000, 50000, and so on. In other words, $f(x)$ is strictly postivie in its domain in this case, as well.}
\label{f when M<11/18}
\end{figure}
The second contour plot (FIG. \ref{f when M<11/18}) shows a similar feature such that its first contour line corresponds to $f(3\Lambda) = 25,000$, the next line is at 50,000, and so on. One can go on with greater values of $x$, and it shows the similar behaviour. Therefore, we conclude that the function $f(x)$ is strictly positive in its domain, which is sufficient to ensure the unitarity of the VPA in the linearized massive gravity.

\begin{acknowledgments}
It is our great pleasure to thank S. Hofmann, T. Rug, G. Buchalla, A. Franca, and A. Vairo for useful discussions. Sungmin Hwang was supported by International Max Planck Research School (IMPRS) from Max-Planck-Institut f{\"u}r Physik, M{\"u}nchen and Excellence Cluster `Universe', and Dennis Schimmel was supported by CeNS, SFB TR 12, NIM, Quantum Initiative Munich. 
\end{acknowledgments}

%\bibliographystyle{plain}
%\bibliography{bibliography}

\begin{thebibliography}{99}
 \bibitem{HiguchiBound} A.~Higuchi, Nucl. Phys, B 282, 397-436, (1987)
 \bibitem{DarkE1} S.~Perlmutter, et al., Atrophys, J. 517, 565 (Supernova Cosmology Project), (1999)
 \bibitem{DarkE2} A.~Reiss, et al., Astro. J. 116, 1009 (Supernova Search Team), (1998)
 \bibitem{WeinbergCC} S.~Weinberg, Rev. Mod. Phys, 61, No.1, (1989)
 \bibitem{Stuckelberg} E.~C.~G. St{\"u}ckelberg, Helv. Phys. Act. 30, 209 (1957)
 \bibitem{Gmass} A.~S.~Goldhaber, M.~M.~Nieto, Rev. Mod. Phys, 82, 939 (2010)
 \bibitem{Gmass2} A.~Gruzinov, New Astro, 10, 311-314, (2005)
 \bibitem{Gmass3} S.~L.~Finn, P.~J.~Sutton, Phys. Rev. D 65, 044022 (2002)
 \bibitem{GiaStefan} G.~Dvali, S.~Hofmann, J.~Khoury, Phys. Rev. D 76, 084006, (2007)
 \bibitem{FPmass} M.~Fierz, W.~Pauli, Proc. R. Soc. A 173, 211, (1939)
 \bibitem{ParkerToms} L.~Parker, D.~Toms ``Quantum Field Theory in Curved Spacetime", Cambridge, UK: Univ. Pr. (2009)
 \bibitem{GibbonsHawking} G.~W.~Gibbons, S.~W.~Hawking, Phys. Rev. D, 15, 10, (1977)
 \bibitem{BH} M.~Banados, Phys. Rev. D 57, No.2, (1998)
 \bibitem{Berkhahn} F.~Berkhahn, D.~Dietrich, S.~Hofmann, JCAP11, 018 (2010)
 \bibitem{Alex} S.~Folkerts, A.~Pritzel, N.~Wintergerst, LMU-ASC 30/11 (2011) [arXiv:1107.3157v2 [hep-th]]
 \bibitem{Birell} N.~D.~Birell, P.~C.~W.~Davies ``Quantum fields in curved space", Cambridge, UK: Univ. Pr. (1982)
 \bibitem{Mukhanov} V.~F.~Mukhanov, S.~Winitzki, ``Introduction to Quantum Effects in Gravity", Cambridge, UK: Univ. Pr. (2007)
 \bibitem{vDVZ1} H.~van Dam, M.~J.~G.~Veltman, Nucl. Phys. B22, 397 (1970)
 \bibitem{vDVZ2} V.~I.~Zakarov, JETP Lett. 12.312 (1970)
 \bibitem{Vainshtein} A.~Vainshtein, Phys. Lett. B39, No.3 (1972)
 \bibitem{Wald} R.~Wald, ``General Relativity", Chicago, USA: Univ. Pr. (1984)
 \bibitem{WeinbergQFT} S.~Weinberg, ``Quantum Theory of Fields" Vol. I, Campridge, UK: Univ. Pr. (1995)
 \bibitem{Zee} A.~Zee, ``Quantum Field Theory in a Nutshell", Princeton, USA: Univ. Pr. (2003)
 \bibitem{Liddle} A.~Liddle ``An Introduction to Modern Cosmology'' (2nd ed.), Chichester; Hoboken, NJ: Wiley (2007)
 \bibitem{Stompor} R.~Stompor et al. ApJ. 561, L7 (2001)
 \bibitem{Spergel} D.~N.~Spergel et al. ApJS. 170, 377 (2007)
 \bibitem{Aragone} C.~Aragone, S.~Deser, Nuov. Cim. 3A, 709 (1971); Nuov. Cim. 57B, 33 (1980)
 \bibitem{Neupane} I.~Neupane, Class. Quantum Grav. 19: 1167-1184 (2002)
 
\end{thebibliography}

\end{document}